\documentclass[pdflatex,sn-mathphys-num]{sn-jnl}

\usepackage{graphicx}%
\usepackage{multirow}%
\usepackage{amsmath,amssymb,amsfonts}%
\usepackage{amsthm}%
\usepackage{mathrsfs}%
\usepackage[title]{appendix}%
\usepackage{xcolor}%
\usepackage{textcomp}%
\usepackage{manyfoot}%
\usepackage{booktabs}%
\usepackage{algorithm}%
\usepackage{algorithmicx}%
\usepackage{algpseudocode}%
\usepackage{listings}%

\theoremstyle{thmstyleone}%
\theoremstyle{thmstyletwo}%

\theoremstyle{thmstylethree}%

\usepackage{lineno}
\usepackage{dsfont}

\begin{document}
\newcommand{\TFJ}[1]{{\color{red}#1}}
\newcommand{\XFJ}[1]{{\color{blue}#1}}

\title[Article Title]{NMRPeak: a ready-to-use intelligent system for molecular structure elucidation enabled by synergistic cross-modal learning}


\author[1,2,3]{\fnm{Fanjie} \sur{Xu}}
\author[2,4]{\fnm{Jinyuan} \sur{Hu}}
\author[4]{\fnm{Jingxiang} \sur{Zou}}
\author[3,5]{\fnm{Junjie} \sur{Wang}}
\author[6]{\fnm{Boying} \sur{Huang}}
\author[3,7]{\fnm{Zhifeng} \sur{Gao}}
\author*[3,7]{\fnm{Xiaohong} \sur{Ji}}\email{jixh@dp.tech}
\author[7,8,9]{\fnm{Weinan} \sur{E}}
\author[2,4]{\fnm{Zhong-Qun} \sur{Tian}}
\author*[6,1,4]{\fnm{Fujie} \sur{Tang}}\email{tangfujie@xmu.edu.cn}
\author*[2,1,4]{\fnm{Jun} \sur{Cheng}}\email{chengjun@xmu.edu.cn}

\affil*[1]{\orgdiv{Institute of Artificial Intelligence}, \orgname{Xiamen University}, \orgaddress{\city{Xiamen}, \postcode{361005}, \country{China}}}

\affil[2]{\orgdiv{State Key Laboratory of Physical Chemistry of Solid Surfaces, iChEM, College of Chemistry and Chemical Engineering}, \orgname{Xiamen University}, \orgaddress{\city{Xiamen}, \postcode{361005}, \country{China}}}

\affil[3]{\orgdiv{DP Technology}, \orgaddress{\city{Beijing}, \postcode{100080}, \country{China}}}

\affil[4]{\orgdiv{Laboratory of AI for Electrochemistry (AI4EC), Tan Kah Kee Innovation Laboratory (IKKEM)}, \orgaddress{\city{Xiamen}, \postcode{361005}, \country{China}}}

\affil[5]{\orgdiv{College of Chemistry and Molecular Engineering, Peking University}, \orgaddress{\city{Beijing}, \postcode{100871}, \country{China}}}

\affil[6]{\orgdiv{Pen-Tung Sah Institute of Micro-Nano Science and Technology, Discipline of Intelligent Instrument and Equipment, iChEM}, \orgname{Xiamen University}, \orgaddress{\city{Xiamen}, \postcode{361005}, \country{China}}}

\affil[7]{\orgname{AI for Science Institute}, \orgaddress{\city{Beijing}, \postcode{100080}, \country{China}}}

\affil[8]{\orgdiv{Center for Machine Learning Research}, \orgname{Peking University}, \orgaddress{ \city{Beijing}, \postcode{100871}, \country{China}}}

\affil[9]{\orgdiv{School of Mathematical Sciences}, \orgname{Peking University}, \orgaddress{ \city{Beijing}, \postcode{100871}, \country{China}}}



\abstract{
One-dimensional nuclear magnetic resonance (NMR) spectroscopy is essential for molecular structure elucidation in organic synthesis, drug discovery, natural product characterization, and metabolomics, yet its interpretation remains heavily dependent on expert knowledge and difficult to scale. Although machine learning has been applied to NMR spectrum prediction, library retrieval, and structure generation, these tasks have largely evolved in isolation using simulated data and incompatible spectral representations, limiting their utility in real experimental settings.
Here we present NMRPeak, a unified cross-modal learning system that integrates these three tasks through an experimentally grounded design. We curate approximately 1.8 million experimental and simulated spectra to construct the largest benchmark for NMR-based structure elucidation. Systematic benchmarking reveals that existing models trained predominantly on simulated spectra suffer severe performance degradation on experimental data, with top-1 structure elucidation accuracy falling below 20\%, underscoring a critical gap between current simulation paradigms and practical deployment.
To bridge this gap, we introduce NMRPeak, a unified system built on a molecule-to-spectrum paradigm with synergistic coupling of prediction, retrieval, and generation modules. The framework is underpinned by a chemically-aware adaptive tokenizer that dynamically balances discretization granularity to preserve spectral semantics while controlling vocabulary size, and a peak-aware similarity metric that enables direct comparison across spectra without requiring peak assignments.
NMRPeak overcomes the longstanding simulation-to-experiment gap in spectrum prediction while achieving over 95\% top-1 accuracy in molecular retrieval and approximately 75\% top-1 accuracy in stereochemistry-aware \textit{de novo} structure generation, substantially outperforming prior methods under experimental conditions. These capabilities establish a foundation for automated and high-throughput molecular structure elucidation in organic synthesis, drug discovery, and chemical biology, positioning NMRPeak as a robust and ready-to-use platform for AI-assisted NMR analysis.
}




\maketitle

\section{Introduction}\label{Introduction}

\par One-dimensional nuclear magnetic resonance (NMR) spectroscopy provides fingerprint-level information that reflects the local chemical environments and connectivity of atoms in organic molecules~\cite{hu2023machine, george2025structure}. As a result, it remains one of the most widely used experimental techniques for molecular structure elucidation~\cite{hartman2016benchmark, lin2021unravelling, lin2022combining, lin2022machine, atwi2022automated, you2025decoding, fan2026relative}. However, conventional NMR interpretation relies heavily on expert knowledge and manual reasoning, making the process time-consuming, labor-intensive, and difficult to scale to complex structures or high-throughput experimental pipelines~\cite{xue2023advances, kuhn2024nuclear, das2025exploring, luo2025deep}. Artificial intelligence (AI) methods have been increasingly applied to address these challenges in NMR-based analysis. Conceptually, NMR interpretation can be formulated as a pair of complementary problems: the forward problem predicts NMR spectra from molecular structures, while the inverse problem infers molecular structures from observed spectra. In practice, AI-assisted NMR analysis has evolved into three major task paradigms: {\textbf{Prediction} models learn to simulate spectra from molecular structures, enabling rapid computational screening and spectrum validation; \textbf{Retrieval} methods search for candidate molecules in databases given a query spectrum, leveraging pre-existing chemical knowledge for efficient structure identification; and \textbf{Generation} approaches attempt to directly construct molecular structures from spectral inputs, enabling \textit{de novo} structure elucidation when database matches are unavailable.

However, prior efforts in AI-assisted NMR analysis have largely focused on specific isolated tasks. In spectrum prediction, Kuhn et al. leveraged graph neural networks (GNNs) to map atomic environments directly to chemical shifts~\cite{jonas2019rapid}. Building on this, Xu et al. advanced the field by developing the NMRNet model that introduces a pre-training and fine-tuning paradigm and establishes a unified framework for NMR chemical shift prediction~\cite{xu2025toward}. For molecular retrieval, prevailing approaches rely on molecule–spectrum contrastive learning, which ranks candidates based on the cosine similarity between the query spectrum and learned molecular embeddings~\cite{yang2021cross, sun2024cross, xu2024enhancing}. Alternatively, Jin et al. proposed NMR-Solver~\cite{jin2025nmr}, which employs NMRNet~\cite{xu2025toward} as a workhorse to simulate spectra for candidate molecules, thereby facilitating an indirect spectrum-to-spectrum retrieval approach to circumvent direct cross-modal alignment issues. Regarding \textit{de novo} generation, recent models have employed autoregressive or diffusion-based mechanisms to achieve end-to-end resolution of molecular structures from one-dimensional NMR spectra~\cite{alberts2023learning, alberts2024unraveling, yao2023conditional, xue2025nmrmind, xiong2025atomic, chen2025dise, yang2026diffnmr}. Furthermore, Hu et al. augmented this approach by incorporating substructure prediction as an auxiliary task within a multi-task learning framework, thereby enhancing the model's comprehension of structural semantics~\cite{hu2024accurate, hu2025pushing}. 

Despite notable progress in each area, existing solutions have largely evolved in isolation, preventing exploitation of their inherent complementarities and limiting their collective impact on real-world structure elucidation. Moreover, several fundamental challenges limit their applicability in practical applications~\cite{chen2025dise, yang2026diffnmr}.

First, spectral representations are misaligned across tasks. Prediction models typically rely on atom-level assignments that map individual atoms to chemical shifts~\cite{paruzzo2018chemical, jonas2019rapid, guan2021real, zou2023deep, ai2024very, yan2025general, bhadauria2025cascade, xu2025toward}, requiring fully annotated training data that are rarely available in experimental settings. In contrast, retrieval and generation methods often operate on unassigned experimental spectra represented as global peak sets, where equivalent atoms are merged and additional information such as splitting patterns, coupling constants and peak integrals in $^{1}$H NMR is present. This representational divide creates a critical gap: atom–shift representations, while computationally convenient, cannot be directly applied to real experimental spectra, whereas molecule–spectrum representations naturally align with how spectra are acquired and reported in practice. Compounding this challenge, spectral discretization strategies remain inconsistent and fundamentally unresolved. Continuous spectral data points  preserve raw spectral information but are highly sensitive to preprocessing choices such as interpolation, resolution, phase correction, and truncation. Discrete peak list representations align with experimental reporting but introduce a critical trade-off in discretization granularity during model tokenization. Fine-grained binning (e.g., 0.1 ppm resolution for $^{13}$C NMR shifts yielding 2,000 bins in Wang et al.~\cite{yao2023conditional, xue2025nmrmind}) leads to large vocabularies and severe data sparsity. Conversely, coarse-grained binning (e.g., fewer than 100 bins in Hu et al.~\cite{hu2024accurate, hu2025pushing}) collapses distinct spectra into nearly identical representations, obscuring chemically meaningful differences. This divergence highlights the absence of a principled, chemically informed solution to the representation problem.

Second, the scarcity of large-scale, curated experimental benchmarks has forced most existing methods to rely on simulated data~\cite{alberts2023learning, alberts2024unraveling, xue2025nmrmind, chen2025dise, yang2026diffnmr}. Although simulation-based datasets enable data scaling, the distribution mismatch between simulated and experimental spectra leads to severe performance degradation when models trained on simulated data are deployed in real-world scenarios. Recent efforts leveraging large language models to extract experimental spectra from literature and patents, such as NMRexp~\cite{wang2025nmrexp} and NMRBank~\cite{wang2025nmrextractor}, partially alleviate this shortage. However, automated extraction pipelines based on large language models often retain various forms of noise (including annotation errors, inconsistent formatting, and incomplete records), resulting in degraded model performance when used directly for training~\cite{shen2025molspectllm}. Systematic curation is therefore required before these resources can serve as reliable benchmarks. Critically, the extent of performance degradation when models trained on simulated data are deployed on experimental spectra has not been systematically quantified, leaving the field without clear guidance on the relative importance of data scale versus data domain.

To address these challenges, we introduce NMRPeak, a unified cross-modal learning framework designed to simultaneously solve NMR spectrum prediction, molecular retrieval and structure generation. We curate approximately 1.8 million experimental and simulated spectra to construct the largest benchmark for NMR-based structure elucidation. This benchmark enables a systematic quantification of the simulation-to-experiment performance degradation. We introduce a chemically-aware adaptive tokenizer that dynamically balances discretization granularity to preserve spectral semantics while controlling vocabulary size, and an assignment-free peak-aware similarity metric that enables direct comparison between predicted and experimental spectra. By tightly coupling prediction, retrieval, and generation through a shared representation and evaluation framework, NMRPeak marks a significant advance on experimental benchmarks by resolving the simulation-to-experiment discrepancy in spectrum prediction and achieving over 95\% top-1 accuracy in molecular retrieval and approximately 75\% top-1 accuracy in stereochemistry-aware \textit{de novo} generation. As such, our approach establishes strong cross-task synergy and achieves substantial performance gains on real-world experimental NMR structure elucidation, enabling a meaningful paradigm shift for both forward and inverse spectral interpretation tasks.

\section{Results}\label{Results}
\subsection{Overview of the NMRPeak system}

As illustrated in Fig.~\ref{fig:framework}a, the NMRPeak system integrates three core modules: the prediction module (NMRPeak-P), the retrieval module (NMRPeak-R), and the generation module (NMRPeak-G), to address diverse scenarios in NMR-based structure elucidation. NMRPeak-P adopts a global molecule-to-spectrum perspective to predict complete NMR spectra from molecular structures, effectively bridging forward prediction and inverse reasoning through unified spectral representation. NMRPeak-R leverages a contrastive learning framework to enable bidirectional cross-modal retrieval, supporting both molecule-to-spectrum and spectrum-to-molecule queries. NMRPeak-G focuses on end-to-end inference of candidate molecular structures directly from experimental NMR spectra, including full stereochemical resolution. A web-based tool\footnote{Available at \url{https://ai4ec.ac.cn/apps/nmrpeak}} further enhances accessibility, enabling streamlined chemical shift predictions for the research community.

\begin{figure}[!t]
\centering
\includegraphics[width=0.94\textwidth]{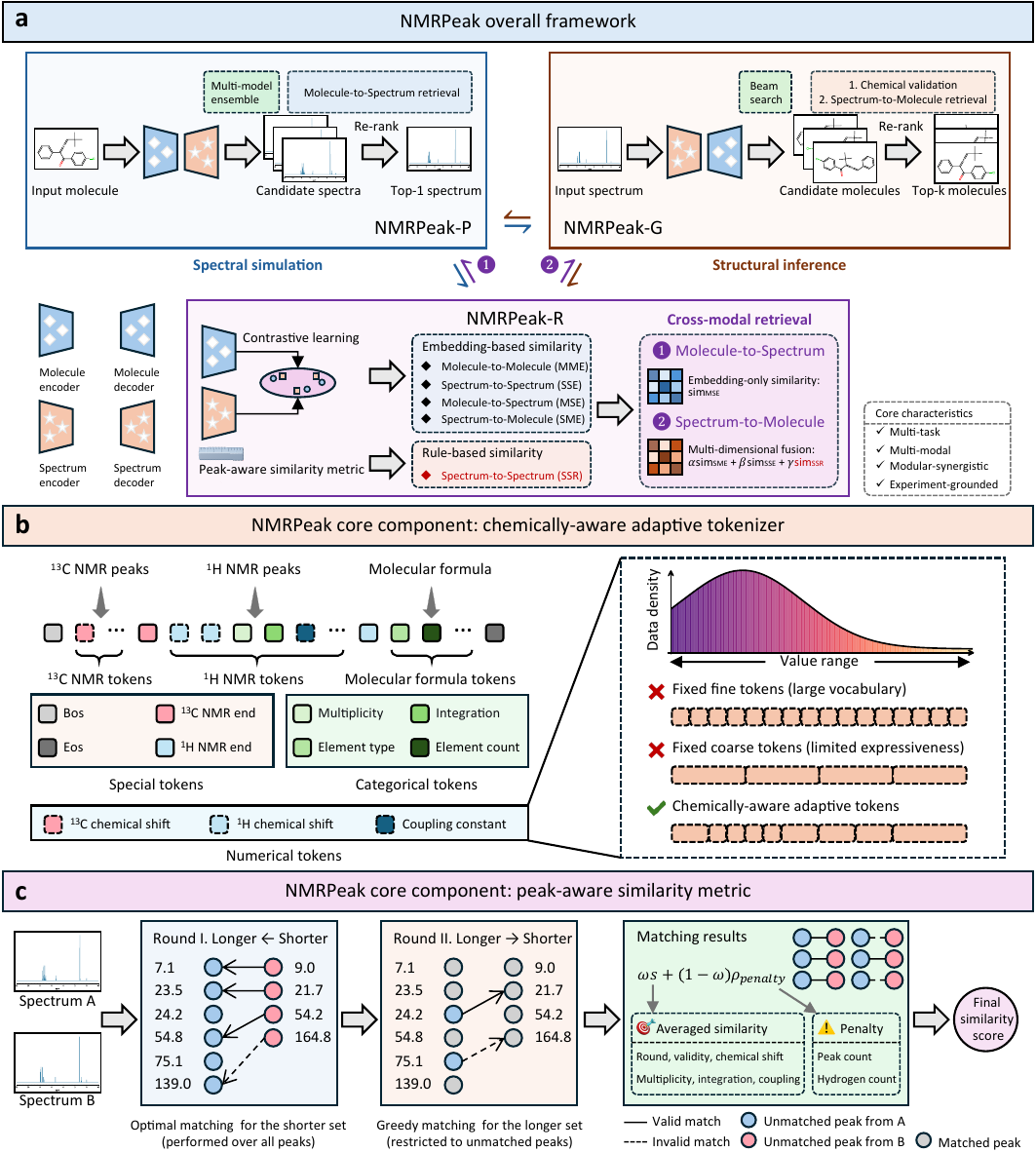}
\caption{\textbf{The NMRPeak framework.} 
\textbf{a}, Overall architecture of NMRPeak, integrating three synergistic modules: NMRPeak-P for forward spectral simulation, and NMRPeak-R for cross-modal retrieval, and NMRPeak-G for inverse structural inference.  
\textbf{b}, The chemically-aware adaptive tokenizer. This component encodes $^{13}$C and $^{1}$H NMR peaks together with molecular formulas into a unified token space that includes special, categorical, and numerical tokens. The adaptive discretization strategy (right) optimizes the trade-off between vocabulary size and semantic resolution by dynamically adjusting token density based on prior knowledge of the data distribution. 
\textbf{c}, The peak-aware similarity metric for assignment-free spectral comparison. The algorithm employs a two-round matching process with count-based penalties to compute a final similarity score.}
\label{fig:framework}
\end{figure}

To facilitate effective cross-modal interaction, we employ specialized encoder architectures tailored to each modality. For understanding of molecular structures, we utilize Uni-Mol~\cite{zhouuni} to encode 3D conformational information, recognizing that 3D geometry strongly influences spectral properties such as through-space coupling and shielding effects. For SMILES generation, we leverage a BART~\cite{lewis2020bart} decoder architecture that supports autoregressive sequence modeling with attention mechanisms. For NMR spectra, we adopt a peak-sequence representation processed through a BART~\cite{lewis2020bart} encoder-decoder architecture for peak-sequence understanding and generation, enabling flexible handling of variable-length spectra and rich spectral attributes including chemical shifts, multiplicities, coupling constants, and peak integrals. This architectural choice naturally accommodates the unordered nature of peak sets and the variable information content across different spectral types. Detailed specifications of these encoder-decoder architectures and their training procedures are provided in the Methods section.

Beyond modular assembly, NMRPeak establishes synergistic interactions that enable mutual enhancement across tasks. NMRPeak-P generates simulated spectra for candidate molecules identified by retrieval or generation, enabling re-ranking based on direct spectrum-to-spectrum similarity rather than relying solely on learned embeddings. This physical validation step substantially improves discrimination between structurally similar candidates. Within NMRPeak-R, molecule-to-spectrum retrieval assists NMRPeak-P by ranking candidate spectra to identify the most accurate simulations, while spectrum-to-molecule retrieval provides structural priors and facilitates candidate re-ranking to accelerate inference in NMRPeak-G. NMRPeak-G assesses the fidelity of NMRPeak-P predictions by testing whether predicted spectra can accurately infer the correct molecular structure. Furthermore, it constructs novel candidate structures \textit{de novo} when the target molecule is absent from reference databases, thereby overcoming the inherent limitations of closed-set retrieval systems.

A key technical obstacle in modeling NMR spectra is spectral discretization. Fixed-width binning schemes struggle to balance sparsity and semantic resolution: fine-grained bins preserve spectral detail but generate sparse, high-dimensional representations prone to overfitting, while coarse-grained bins reduce dimensionality but collapse chemically distinct signals into indistinguishable categories. To resolve this trade-off, we introduce a chemically-aware adaptive tokenizer that dynamically adjusts quantization granularity based on domain-specific chemical knowledge. As illustrated in Fig.~\ref{fig:framework}b, ultra-fine resolution is used in dense fingerprint regions, while coarser resolution is applied in sparse regions. These adaptive strategies are employed for $^{1}$H chemical shifts, $^{13}$C chemical shifts, and $^{1}$H coupling constants. In addition to these adaptive numerical tokens, the tokenizer incorporates standard categorical tokens for elemental and peak-specific attributes, as well as special tokens to manage sequence boundaries. This chemically informed discretization converts variable-length spectra into compact token sequences that preserve information-rich fine structure while controlling vocabulary size, reducing data sparsity in low-signal regions, and improving computational efficiency during training and inference. The complete tokenization scheme, including region boundaries and resolution tables for each numerical token type, is provided in the Methods section.

As illustrated in Fig.~\ref{fig:framework}c, another core component of this integrated framework is the peak-aware similarity metric, which addresses a fundamental challenge in evaluating unassigned experimental spectra. Traditional atom-to-shift metrics require explicit atom-level assignments that are rarely available experimentally in the literature data. Our metric instead quantifies the correspondence between the predicted and experimental peak sets through a robust two-stage bipartite matching procedure. In the first stage, an optimal matching is performed for the shorter peak set to ensure a rigorous global alignment of primary spectral features. This is followed by a second stage of greedy matching for the remaining unmatched peaks in the longer set, specifically designed to tolerate discrepancies such as spurious or missing peaks that frequently occur in real-world measurements. To prevent excessive tolerance for such deviations, we incorporate explicit penalty terms based on inconsistencies in peak count and hydrogen count. This multifaceted evaluation approach not only enables rigorous assessment of molecule-to-spectrum prediction but also provides a quantitative foundation for spectrum-to-spectrum retrieval, which serves as a critical ranking mechanism for both retrieval and generation modules. The mathematical formulation of this metric, including the feature-specific weighting schemes, is detailed in the Methods section.

To address the persistent discrepancy between simulated and experimental NMR data, we construct NMRPeak using large-scale experimental benchmarks rather than relying exclusively on computational simulations. We adopt NMRexp~\cite{wang2025nmrexp} as our primary data source, representing the largest open repository of experimental NMR spectra to date with over three million entries extracted from scientific literature and supporting information through automated text mining and manual curation. To explicitly quantify the distribution shift between data domains, we incorporate the simulated NMR subset from Alberts et al. (MST-NMR)~\cite{alberts2024unraveling}, one of the most widely used simulated NMR datasets in previous work~\cite{alberts2023learning, alberts2024unraveling, xiong2025atomic, yang2026diffnmr}, enabling direct comparison of model performance across simulated and experimental scenarios. Our benchmark comprises approximately 1 million experimental spectra from NMRexp~\cite{wang2025nmrexp} and 0.8 million simulated spectra from MST-NMR~\cite{alberts2024unraveling} for training, validation, and testing. Detailed statistics, including a UMAP~\cite{mcinnes2018umap} visualization of the structural distribution overlap between the simulated and experimental domains and the specific data split across sets, are summarized in Fig.~\ref{fig:results}a-b.

\begin{figure}[!t]
\centering
\includegraphics[width=0.94\textwidth]{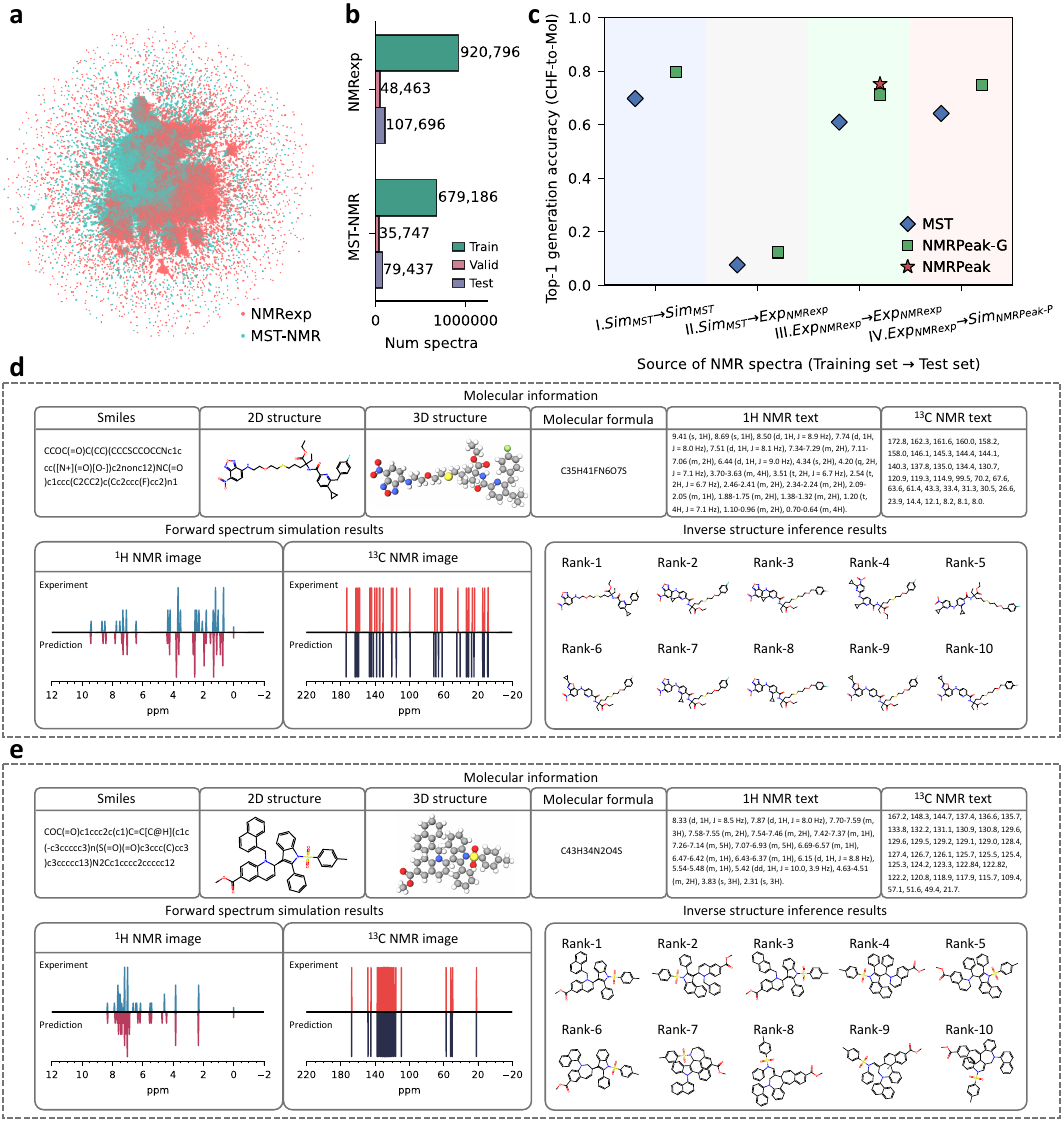}
\caption{\textbf{Data benchmarking and performance evaluation of NMRPeak on experimental datasets.} 
\textbf{a}, UMAP~\cite{mcinnes2018umap} projection of structural distributions between simulated (MST-NMR~\cite{alberts2024unraveling}) and experimental (NMRexp~\cite{wang2025nmrexp}) datasets.
\textbf{b}, Statistical summary of the curated NMR benchmark, comprising over 1.8 million structure–spectrum pairs across training, validation, and test splits for both simulated and experimental domains. 
\textbf{c}, Comparative analysis of top-1 structure elucidation accuracy under different training-to-test scenarios for the baseline MST model~\cite{alberts2024unraveling}, NMRPeak-G (single module), and the unified NMRPeak framework, where each molecule is characterized by $^{13}$C NMR, $^{1}$H NMR, and molecular formula information.
\textbf{d, e}, Representative case studies of complex molecular structure elucidation. Each panel displays the input molecular information (SMILES, 2D/3D structures, molecule formula, and text-based NMR peaks), alongside the \textbf{forward spectrum simulation results} and \textbf{inverse structure inference results}.}
\label{fig:results}
\end{figure}

Fig.~\ref{fig:results}c provides empirical evidence for NMRPeak design through systematic comparisons with the MST baseline~\cite{alberts2024unraveling} in terms of top-1 structure elucidation accuracy across multiple benchmark configurations. At the data level, models trained exclusively on simulated spectra exhibit severe performance degradation when evaluated on experimental test sets (see Fig.~\ref{fig:results}c, Stage I $\rightarrow$ II). This substantial performance degradation quantifies the distribution mismatch between simulated and experimental NMR spectra. Crucially, the construction of experimental training sets enables a qualitative leap in model capability for real-world scenarios (see Fig.~\ref{fig:results}c, Stage II $\rightarrow$ III). At the representation level, NMRPeak-G consistently outperforms the MST baseline under identical data conditions (see Fig.~\ref{fig:results}c, comparing green squares with blue diamonds), demonstrating that improved spectral tokenization yields performance gains complementary to data quality. At the system level, the integration of NMRPeak-P for simulating candidate spectra, followed by re-ranking via NMRPeak-R, further boosts the top-1 structure elucidation accuracy of NMRPeak-G in real-world experimental scenarios (see Fig.~\ref{fig:results}c, Stage III). Moreover, serving as a `secondary validator' for NMRPeak-P, NMRPeak-G achieves higher structure elucidation accuracy when operating on spectra predicted by NMRPeak-P than when using raw experimental spectra (see Fig.~\ref{fig:results}c, Stage III $\rightarrow$ IV). This seemingly counterintuitive result reveals a key insight: high-quality predicted spectra need not replicate experimental noise to be chemically informative. Instead, NMRPeak-P effectively denoises experimental variability while preserving essential structural semantics, thereby producing spectra that are better suited for downstream reasoning. Moreover, to intuitively demonstrate the system's proficiency in handling high-complexity molecular architectures, we selected two representative examples from the NMRexp test set featuring stereochemical complexity, extended chains, polycyclic frameworks, heteroatom-rich structures, and high atom counts (see Fig.~\ref{fig:results}d-e). These case studies highlight NMRPeak's exceptional performance in both high-fidelity forward spectrum simulation and precise reverse structural inference, where the ground-truth structures were successfully identified and ranked at the top-1 position amidst structurally similar candidates. By simultaneously advancing forward spectrum prediction and inverse structure inference, NMRPeak demonstrates that system-level integration is essential for reliable experimental NMR interpretation.

\subsection{NMRPeak-P: Spectrum prediction from molecular structures}

Unlike conventional NMR prediction models that operate at the atom-to-shift level~\cite{jonas2019rapid, xu2025toward, yan2025general}, mapping individual atoms to specific chemical shift values, NMRPeak-P adopts a global molecule-to-spectrum paradigm to directly predict complete, unassigned experimental NMR spectra. This design choice addresses a fundamental practical constraint: atom-level assignments are rarely available for experimental spectra and typically require extensive expert knowledge. By modeling spectra as unordered sets of peaks rather than atom-indexed outputs, we naturally aligned NMRPeak-P with real-world experimental workflows where spectra are reported as lists of peaks with associated attributes (chemical shifts, multiplicities, integrals, and coupling constants) without explicit atom assignments. This representation provides a unified interface compatible with both downstream retrieval and generation tasks, which operate on unassigned spectral data in the same manner.

\begin{figure}[!t]
\centering
\includegraphics[width=0.94\textwidth]{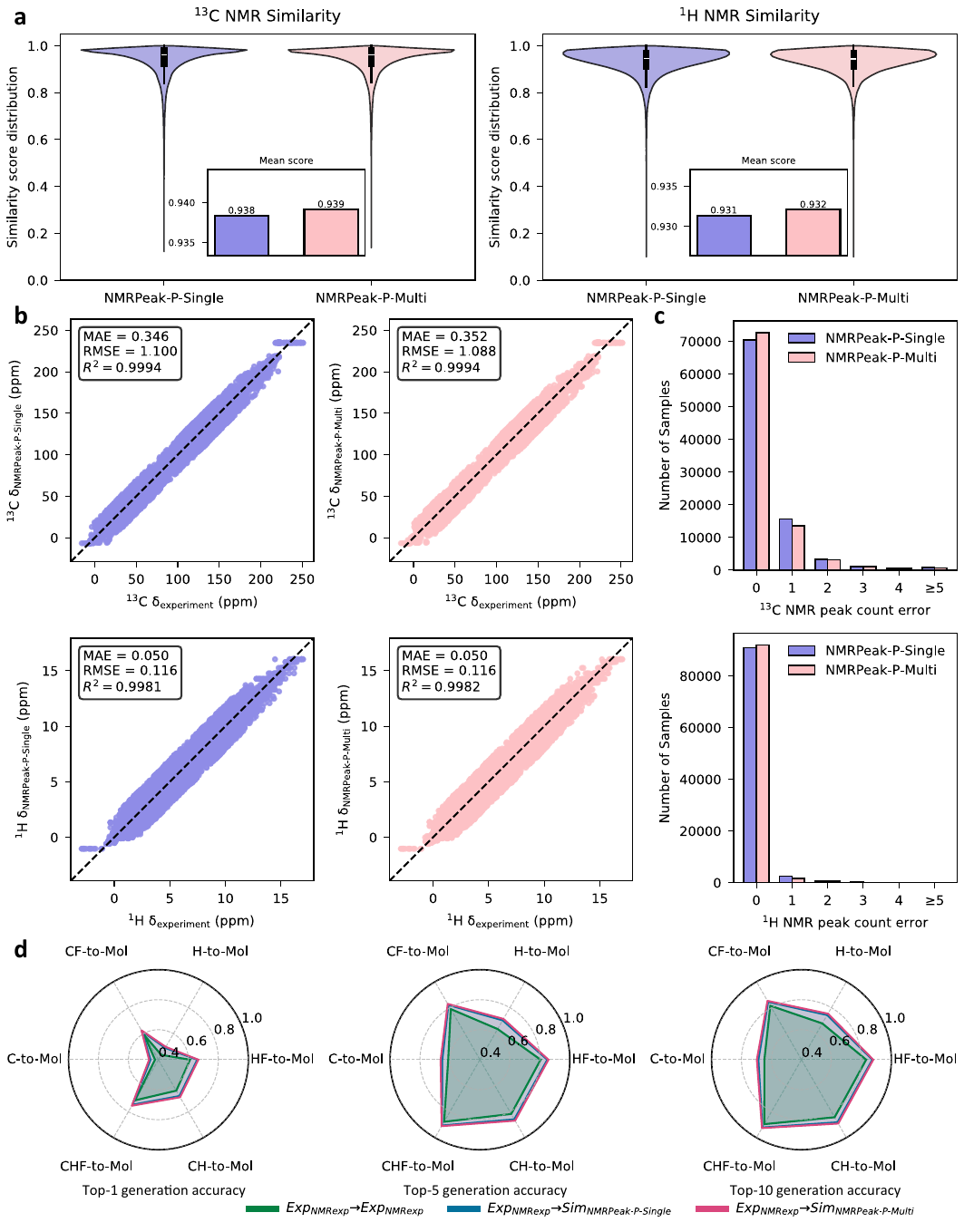}
\caption{\textbf{Performance of NMRPeak-P.} 
\textbf{a}, Distribution of spectral similarity scores between predicted and experimental $^{13}$C (left) and $^{1}$H (right) NMR spectra, calculated using the peak-aware similarity metric. 
\textbf{b}, Correlation between predicted and experimental chemical shifts ($\delta$) for $^{13}$C (top) and $^{1}$H (bottom), derived from the first-round valid matching using the peak-aware similarity metric. 
\textbf{c}, Error distribution of predicted peak counts for $^{13}$C and $^{1}$H NMR spectra. For $^{1}$H spectra, peak counts are expanded based on integration values to ensure physically grounded comparisons.
\textbf{d}, Radar charts show the top-$k$ ($k=1, 5, 10$) molecular generation accuracy of NMRPeak-G using simulated and experimental spectra across six input configurations. For all panels, C and H correspond to $^{13}$C and $^{1}$H NMR peaks, and F denotes molecular formula constraints.}
\label{fig:prediction_results}
\end{figure}

NMRPeak-P generates NMR spectra through autoregressive prediction of discrete spectral tokens produced by the chemically-aware adaptive tokenizer. This formulation allows the model to handle variable-length peak sets and encode multiple spectral attributes beyond chemical shifts, including splitting patterns, coupling constants, and integrals in $^{1}$H NMR. To further improve prediction robustness and accuracy, we implement a multi-model ensemble strategy in which multiple independent predictors generate candidate spectra that are subsequently re-ranked using the molecule-to-spectrum ranking function of NMRPeak-R. This ensemble-based refinement, designated as NMRPeak-P-Multi, effectively identifies predictions that best capture the underlying molecular structure's spectral signature, while the single-model baseline is denoted as NMRPeak-P-Single.

The evaluation of molecule-to-spectrum prediction poses a challenge in the absence of atom-level assignments. Traditional evaluation metrics based on chemical shift differences per-atom are not applicable in this setting. We therefore employ the proposed peak-aware similarity metric, which quantifies the global correspondence between predicted and experimental peak sets using optimal bipartite matching. The metric jointly considers chemical shift accuracy and peak count consistency, and imposes explicit penalties for spurious or missing peaks. 

Fig.~\ref{fig:prediction_results} summarizes the performance of NMRPeak-P on the NMRexp test set. As shown in Fig.~\ref{fig:prediction_results}a, NMRPeak-P-Multi consistently outperforms the single-model variant in both $^{13}$C and $^{1}$H NMR spectra in terms of peak-aware similarity. Although improvements in similarity scores appear modest, a more detailed analysis reveals substantial gains in structurally meaningful aspects of the predictions. For peaks successfully matched in the first matching stage, both single and ensemble models achieve high chemical shift accuracy (see Fig.~\ref{fig:prediction_results}b), indicating that the core spectral features are well captured. In contrast, the ensemble model exhibits a markedly more concentrated error distribution in peak count prediction (see Fig.~\ref{fig:prediction_results}c), with the majority of molecules showing deviations of fewer than five peaks.

Beyond direct spectral similarity, we assess the functional quality of predicted spectra by examining their utility for inverse structure inference. Specifically, we apply the frozen NMRPeak-G to spectra predicted by NMRPeak-P and measure structure elucidation accuracy. As shown in Fig.~\ref{fig:prediction_results}d, the spectra generated by NMRPeak-P enable higher structure inference accuracy than the raw experimental spectra under identical conditions. When $^{13}$C NMR, $^{1}$H NMR and molecular formula information are provided, NMRPeak-G achieves a top-1 accuracy of 71.29\% using experimental spectra. This accuracy increases to 74.73\% for spectra generated by NMRPeak-P-Single and further improves to 75.42\% for spectra from NMRPeak-P-Multi. Collectively, these results underscore that NMRPeak-P not only bridges the long-standing gap between the simulated and experimental spectra, but also establishes a critical foundation for synergistic interaction between forward prediction and inverse structure elucidation.

\subsection{NMRPeak-R: Cross-modal retrieval between molecules and NMR spectra}

Drawing inspiration from CLIP-style contrastive learning frameworks for cross-modal alignment~\cite{radford2021learning, lai2025end, li2025powder, rocabert2025multi}, NMRPeak-R aligns molecular structures and NMR spectra into a shared latent embedding space to enable bidirectional molecule-to-spectrum and spectrum-to-molecule cross-modal retrieval. The learned joint embedding space provides an efficient mechanism for rapid candidate recall across large-scale repositories, enabling screening of millions of molecule-spectrum pairs through simple cosine similarity computation in the latent space. However, relying exclusively on latent embeddings for spectrum-to-molecule faces fundamental limitations in NMR-based structure elucidation. Specifically, NMR spectra corresponding to structurally similar molecules often exhibit subtle yet chemically decisive differences. These fine-grained distinctions, such as small chemical shift variations or splitting patterns, are difficult to fully capture in a single global embedding vector optimized through contrastive learning objectives. Consequently, embedding-based retrieval alone struggles to discriminate between hard negative candidates with high structural similarity that cluster closely in the latent space despite representing chemically distinct molecules. This limitation becomes increasingly severe as the database size grows and the density of the chemical space increases, creating a performance bottleneck that cannot be overcome through embedding space refinement alone.

\begin{figure}[!t]
\centering
\includegraphics[width=0.94\textwidth]{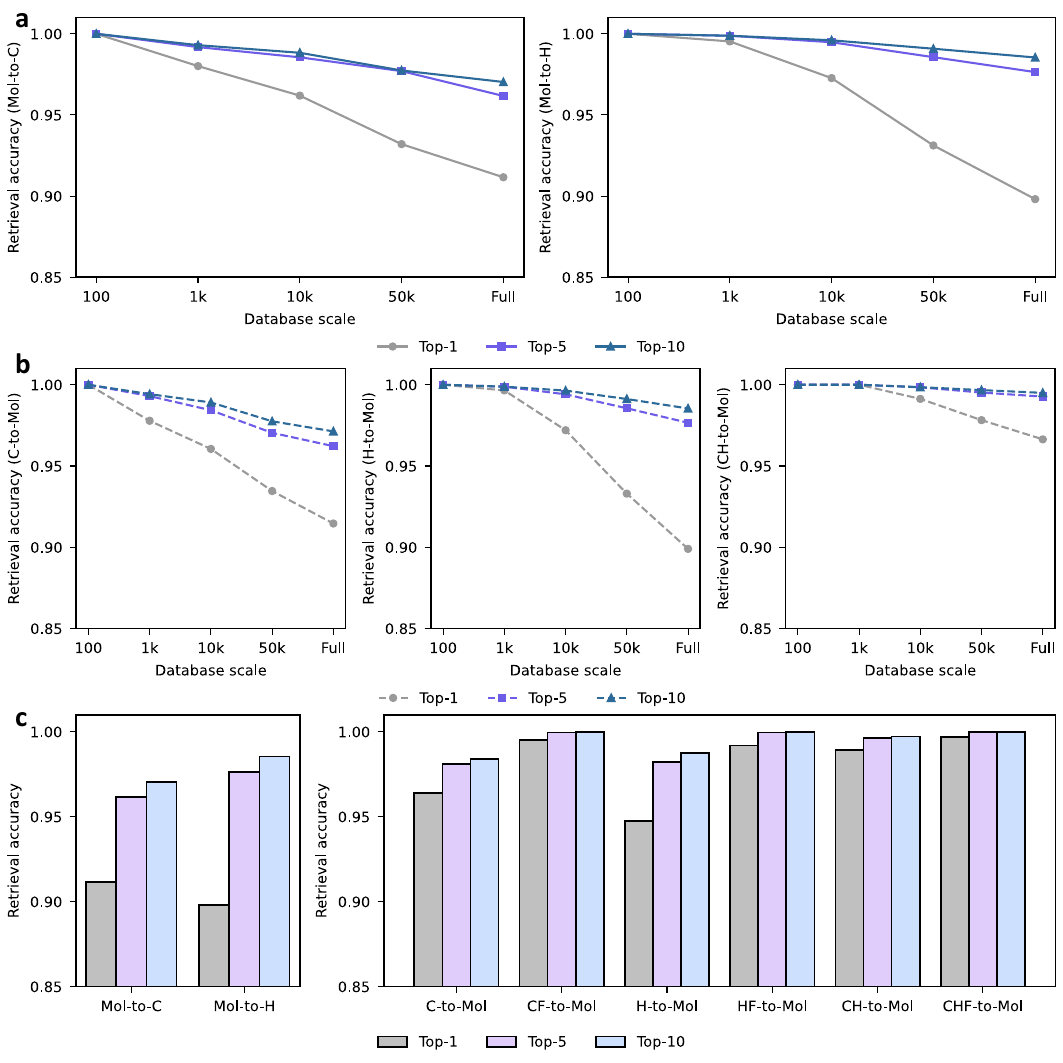}
\caption{\textbf{Performance of NMRPeak-R.} 
\textbf{a, b}, Impact of database scale on retrieval accuracy for molecule-to-spectrum (a) and spectrum-to-molecule (b) tasks using a single contrastive learning strategy. Results are reported for database sizes ranging from 100 entries to the full experimental set ($\approx$100k entries). 
\textbf{c}, Comprehensive retrieval performance on the full-scale benchmark across diverse input modalities. The left panel shows contrastive learning-based molecule-to-spectrum retrieval and the right panel displays spectrum-to-molecule retrieval results using multi-dimensional fusion strategy across six input configurations. For all panels, C and H correspond to $^{13}$C and $^{1}$H NMR peaks, and F denotes molecular formula constraints.}
\label{fig:retrieval_results}
\end{figure}

To address this challenge, NMRPeak-R introduces a multi-dimensional retrieval framework that augments semantic embedding similarity with explicit spectrum-level physical validation in spectrum-to-molecule retrieval. We leverage NMRPeak-P as a bridging mechanism to generate simulated spectra for candidate molecules, thereby enabling direct spectrum-to–spectrum comparison. This design yields three complementary similarity scores: (1) Spectrum-to-molecule embedding similarity (NMRPeak-R-S2M-SME) computes the cosine similarity between the query spectrum embedding and the candidate molecule embedding in the shared contrastive latent space, providing efficient coarse-grained semantic matching; (2) Spectrum-to-spectrum embedding similarity (NMRPeak-R-S2M-SSE) computes the cosine similarity between the latent representation of the query spectrum and the latent representation of the predicted spectrum for each candidate molecule, enabling comparison in the more specialized spectral embedding subspace; (3) Spectrum-to-spectrum rule-based similarity (NMRPeak-R-S2M-SSR) applies the peak-aware similarity metric to directly compare the query spectrum and predicted spectrum in the original signal space. We combine these three complementary scores through a weighted linear combination, designated as NMRPeak-R-S2M-Combine, allowing for rapid candidate retrieval and accurate fine-grained discrimination.

Fig.~\ref{fig:retrieval_results} summarizes the cross-modal retrieval performance of NMRPeak-R after contrastive learning. Fig.~\ref{fig:retrieval_results}a-b demonstrate the scale-dependent behavior of a single contrastive learning strategy across database sizes ranging from 100 molecule-spectrum pairs to the full set (approximately 100,000 entries). The baseline approach exhibits a clear decline in retrieval accuracy as database size increases, with the top-1 accuracy dropping from 100\% at 100 entries to below approximately 90\% at 100,000 entries. This degradation directly reflects the rising prevalence of hard negatives: as the database grows, the probability increases that structurally similar molecules with nearly identical embedding representations will be present, making discrimination based solely on latent similarity increasingly difficult. By incorporating simulated spectra and explicit peak-level comparison, NMRPeak-R significantly mitigates this degradation. Quantitative results in Fig.~\ref{fig:retrieval_results}c demonstrate that the combined multi-dimensional strategy elevates spectrum-to-molecule retrieval accuracy beyond 95\%, even in large-scale settings. When molecular formula constraints are additionally provided, the retrieval accuracy approaches nearly 100\%, underscoring the robustness and practical reliability of the system.

\begin{figure}[!t]
\centering
\includegraphics[width=0.94\textwidth]{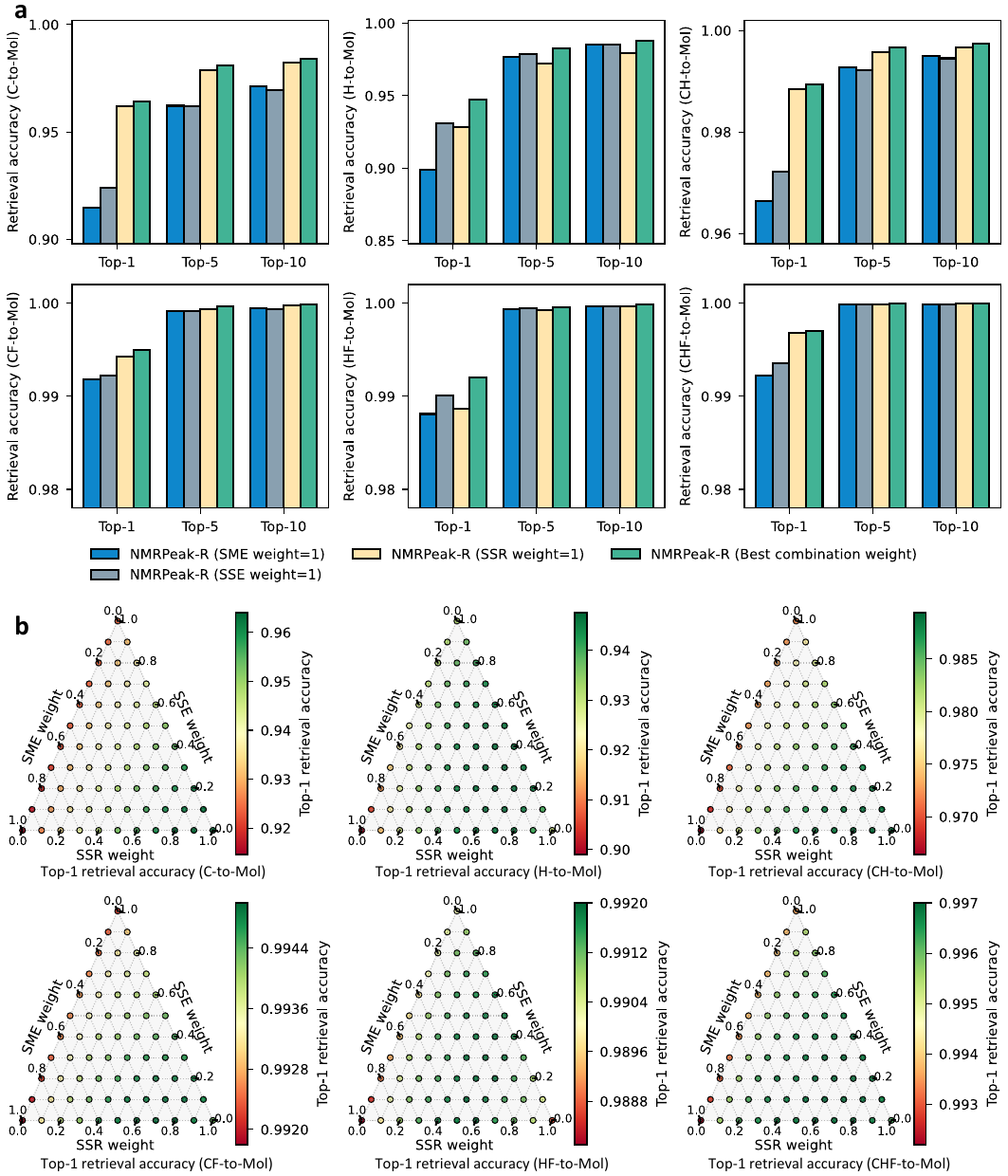}
\caption{\textbf{Ablation study and weighting analysis of NMRPeak-R.} 
\textbf{a}, Comparison of retrieval accuracies for different variants across various input modalities.
\textbf{b}, Ternary plots illustrating the impact of weighting strategies on top-1 retrieval accuracy for diverse configurations. For all panels, SME, SSE, and SSR denote spectrum-to-molecule embedding, spectrum-to-spectrum embedding, and spectrum-to-spectrum rule-based similarity, respectively. C and H correspond to $^{13}$C and $^{1}$H NMR peaks, and F denotes molecular formula constraints.}
\label{fig:retrieval_analysis}
\end{figure}


To dissect the contribution of each component, Fig.~\ref{fig:retrieval_analysis}a compares multiple retrieval variants. Introducing NMRPeak-R-S2M-SSE yields measurable improvements over NMRPeak-R-S2M-SME, confirming the utility of predicted spectra as intermediate representations. Incorporating NMRPeak-R-S2M-SSR leads to a more pronounced performance gain, closely approaching that of the full combined model. These results highlight the importance of directly modeling spectral physics rather than relying exclusively on latent semantic alignment. Further analysis of weighting strategies (see Fig.~\ref{fig:retrieval_analysis}b and Supplementary Figure 1) reveals a consistent pattern: overemphasizing embedding-based similarity degrades retrieval performance, whereas increasing the contribution of explicit peak-aware rules steadily improves accuracy. This observation suggests a general retrieval principle in scientific domains that latent embeddings are effective for candidate recall while explicit domain rules are essential for fine-grained discrimination.

NMRPeak-R proves to be a robust and scalable retrieval module that integrates semantic alignment with physically grounded validation. By tightly coupling prediction and retrieval through simulated spectra and peak-aware similarity, NMRPeak-R forms a critical link in the integrated NMRPeak system. The retrieval module not only enables efficient database search for structure identification, but also provides essential ranking mechanisms that enhance both the prediction module (by selecting the most accurate simulated spectra during ensemble refinement) and the generation module (by providing structural priors and candidate validation during \textit{de novo} structure construction). This bidirectional integration exemplifies the synergistic design philosophy underlying the NMRPeak framework, where each component contributes to and benefits from the capabilities of the others.

\subsection{NMRPeak-G: End-to-end molecular structure generation from NMR spectra}

A central challenge in NMR-based molecular structure elucidation is the inference of complete stereochemical information from spectral data. Many prior studies have avoided this challenge by restricting the output space to two-dimensional molecular graphs or stereochemistry-agnostic representations such as canonical SMILES without stereochemical descriptors~\cite{xiong2025atomic, hu2024accurate, hu2025pushing}. Although such simplifications reduce modeling complexity and improve training convergence, they fail to address the practical requirements of experimental chemists, for whom stereochemical configurations such as chirality and geometric isomerism are often essential. NMRPeak-G addresses this challenge by performing end-to-end molecular generation with full stereochemical resolution. Given an experimental NMR spectrum, the model autoregressively generates a stereochemically complete SMILES sequence, requiring it to infer not only atomic connectivity but also subtle spatial information encoded in spectral features. This task represents a substantially more challenging objective than topology-only reconstruction and places stringent demands on the model's capacity for spectral interpretation.

\begin{figure}[!t]
\centering
\includegraphics[width=0.94\textwidth]{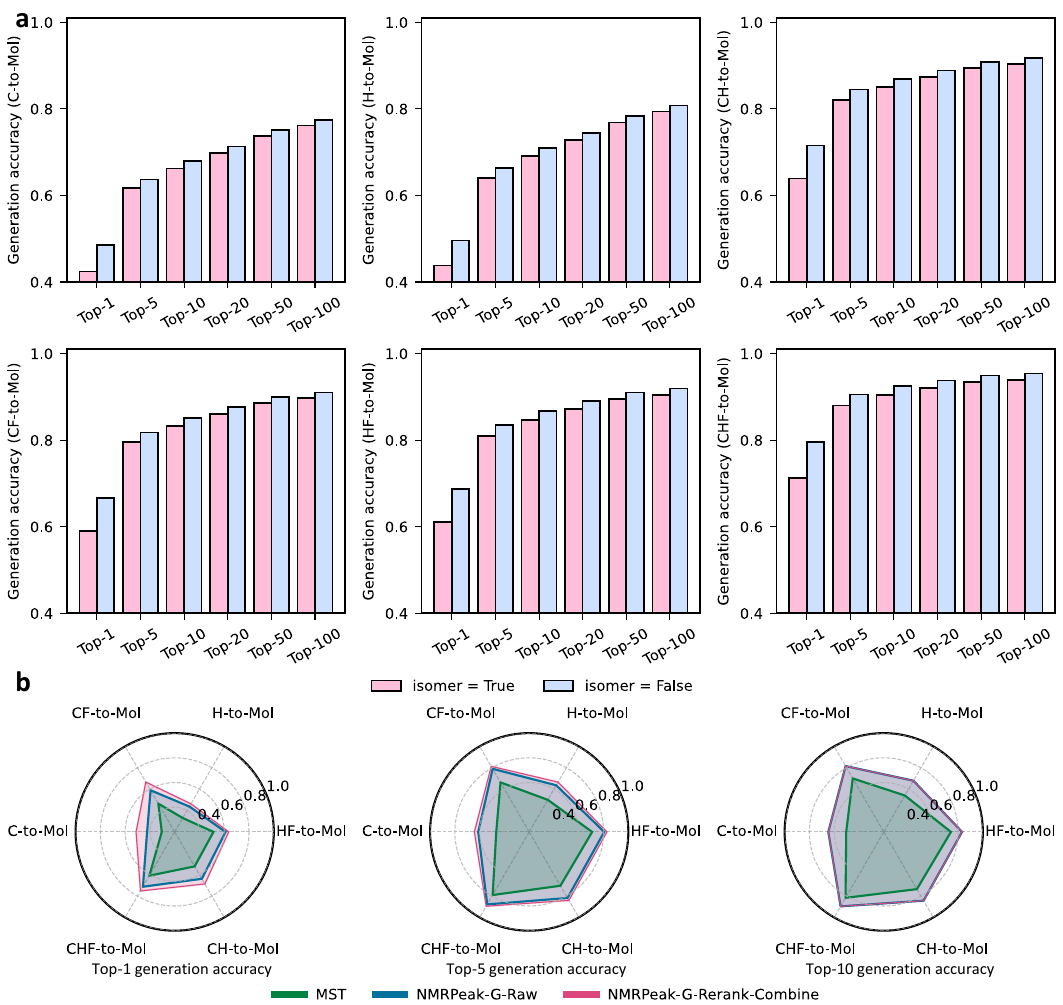}
\caption{\textbf{Performance of NMRPeak-G.} 
\textbf{a}, Impact of beam search width on structure elucidation accuracy across six diverse input configurations. The bar charts compare stereochemistry-aware accuracy (isomer = True) against topology-only accuracy (isomer = False) for top-1 to top-100 candidates. 
\textbf{b}, Comparative evaluation of generative accuracy for the MST baseline, NMRPeak-G-Raw, and the full NMRPeak-G-Rerank-Combine strategy. Radar charts illustrate top-1, top-5, and top-10 generation performance across different input modalities. For all panels, C and H correspond to $^{13}$C and $^{1}$H NMR peaks, and F denotes molecular formula constraints.}
\label{fig:generation_results}
\end{figure}

 The model accepts as input the peak-sequence representation produced by the chemically-aware adaptive tokenizer, which encodes both $^{13}$C NMR and $^{1}$H NMR spectra along with optional molecular formula constraints when available. During inference, the model generates candidate SMILES sequences using beam search with configurable beam widths to explore diverse structural hypotheses. However, autoregressive generation from complex spectral inputs inevitably produces some candidates that contain minor syntactic errors or chemically invalid structures that violate fundamental chemical principles. To address these issues, we employ a progressive refinement strategy that integrates explicit chemical validation and cross-module feedback. 
 
The base generation variant, NMRPeak-G-Raw, produces ranked candidate SMILES sequences directly from beam search scores without post-processing. These candidates are subsequently filtered and re-ranked in NMRPeak-G-Rerank-Base by removing invalid or duplicate SMILES using RDKit~\cite{landrum2025rdkit} validation and by enforcing molecular formula constraints when available to prune the candidate space. To further improve ranking accuracy, we exploit synergistic interactions within the NMRPeak system. Specifically, for each validated candidate molecule, we leverage NMRPeak-P to generate its predicted NMR spectrum and evaluate this simulated spectrum against the original query spectrum using multiple similarity metrics of NMRPeak-R to re-rank candidates. This yields four enhanced variants: NMRPeak-G-Rerank-SME, NMRPeak-G-Rerank-SSE, NMRPeak-G-Rerank-SSR, and NMRPeak-G-Rerank-Combine, mirroring the multi-dimensional re-ranking strategy employed in NMRPeak-R.

\begin{figure}[!t]
\centering
\includegraphics[width=0.94\textwidth]{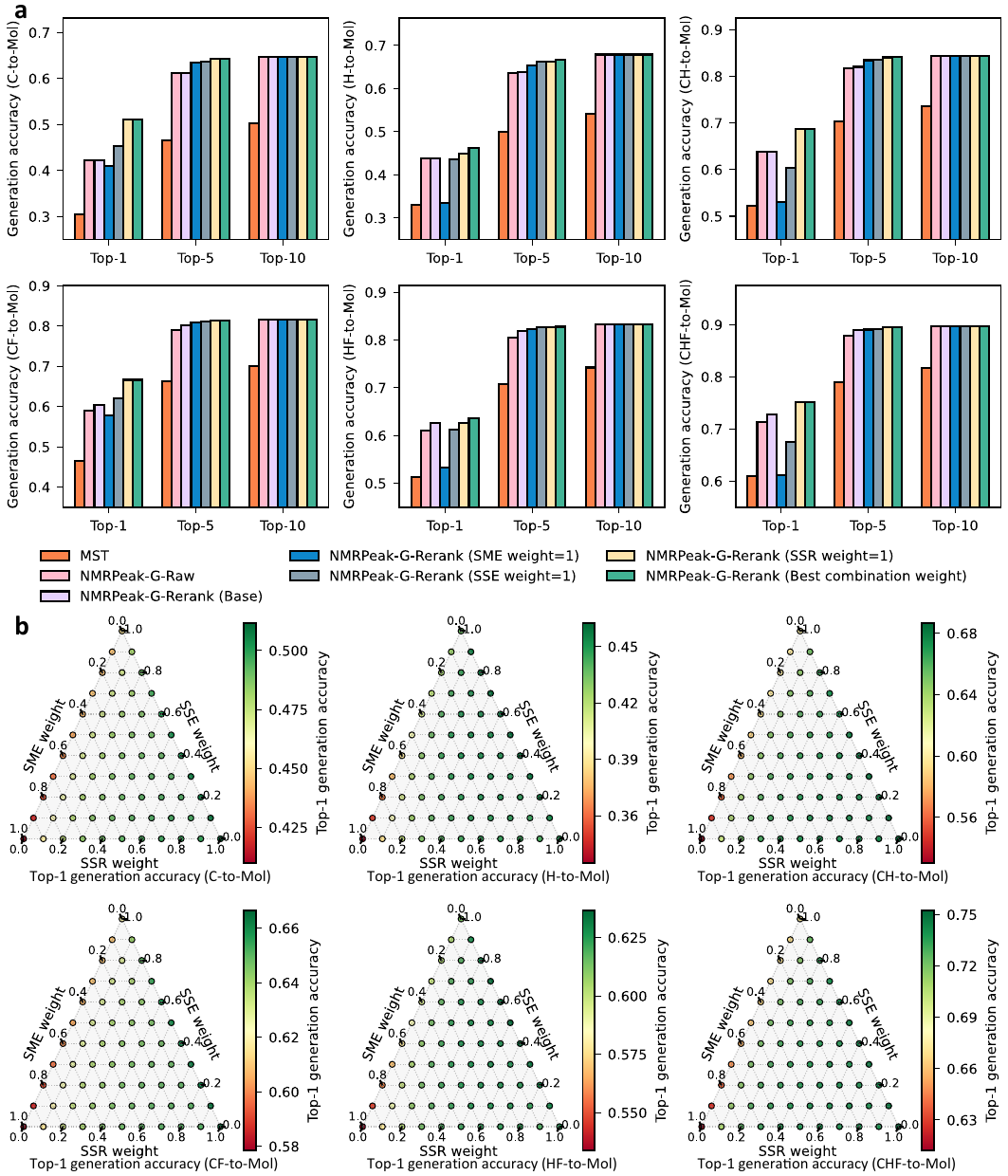}
\caption{\textbf{Ablation study and weighting analysis of NMRPeak-G.} 
\textbf{a}, Comparison of structural inference accuracies for different generative variants across various input modalities. The bar charts evaluate the performance of the MST baseline, NMRPeak-G-Raw, and various re-ranking configurations (Base, SME, SSE, SSR, and their combinations) for top-1, top-5, and top-10 candidates. 
\textbf{b}, Ternary plots illustrating the impact of weighting strategies on top-1 generation accuracy for diverse configurations. For all panels, SME, SSE, and SSR denote spectrum-to-molecule embedding, spectrum-to-spectrum embedding, and spectrum-to-spectrum rule-based similarity, respectively. C and H correspond to $^{13}$C and $^{1}$H NMR peaks, and F denotes molecular formula constraints.}
\label{fig:generation_analysis}
\end{figure}


Fig.~\ref{fig:generation_results}a evaluates structure elucidation accuracy on the NMRexp experimental test set, comparing performance with and without stereochemical consideration under varying beam search widths. Although stereochemistry-aware prediction exhibits lower top-1 accuracy, reflecting the intrinsic difficulty of distinguishing stereoisomers, the performance difference narrows as the beam size increases. Importantly, the stereochemistry-aware accuracy curve converges toward the topology-only upper bound. This behavior demonstrates the model’s capacity to extract and utilize stereochemically relevant information from NMR spectra. Fig.~\ref{fig:generation_results}b compares NMRPeak-G with the MST baseline model under identical experimental conditions and input configurations. In the comprehensive setting where $^{13}$C NMR, $^{1}$H NMR, and molecular formula are all provided, the top-1 accuracy progressively improves from 60.93\% (MST) to 71.29\% (NMRPeak-G-Raw), ultimately reaching 75.22\% with NMRPeak-G-Rerank-Combine. Mirroring this trend across all configurations, NMRPeak-G-Raw achieves substantial performance improvements, demonstrating that the chemically-aware adaptive tokenizer enhances the model's capacity for spectral interpretation and structure inference. The further gains in top-1 accuracy observed upon introducing re-ranking strategies demonstrate that each component of the refinement pipeline contributes meaningfully to final performance.

Ablation studies presented in Fig.~\ref{fig:generation_analysis} and Supplementary Figure 2 further elucidate the contributions of individual components. Specifically, NMRPeak-G-Rerank-Base, by incorporating chemical validation, yields a tangible improvement over NMRPeak-G-Raw baseline, while NMRPeak-G-Rerank-Combine leverages multi-dimensional ranking strategies to further optimize the ranking of candidate molecules. These trends across different ranking components closely mirror the observations from NMRPeak-R, reinforcing a fundamental principle that emerges across both retrieval and generation tasks.

The generation results establish that NMRPeak-G successfully performs end-to-end stereochemistry-aware structure elucidation directly from experimental NMR spectra, achieving accuracy levels that approach practical applications for computer-assisted structure determination. The consistent benefits observed from integrating prediction and retrieval modules for candidate validation underscore the importance of system-level design, demonstrating that isolated task-specific models cannot achieve the same performance as tightly integrated multi-module frameworks where each component enhances the others through complementary capabilities. This synergistic architecture enables NMRPeak-G to overcome the fundamental limitations of closed-set retrieval systems by constructing novel molecular structures \textit{de novo} when target molecules are absent from reference databases, while simultaneously providing validation mechanisms for simulated spectra that improve the accuracy of both forward prediction and database retrieval tasks.

\section{Discussion}\label{Discussion}

The current implementation of NMRPeak establishes a tightly-coupled integration among prediction, retrieval, and generation modules, creating a unified framework where each component enhances the others. This synergistic architecture validates our core premise that separating forward and inverse tasks limits performance, whereas their coupling enables robust interpretation.

Nevertheless, several limitations in the current system suggest critical avenues for future research. First, because of the scarcity of metadata in public repositories, the model does not yet explicitly account for environmental factors such as solvent effects, temperature dependence, or pH sensitivity. Future work must therefore prioritize the curation of richly annotated experimental databases to capture these physicochemical nuances. Second, our reliance on one-dimensional $^{13}$C and $^{1}$H NMR spectra, while effective for topology, may lack the discriminative power for highly complex structures. Extending the framework to incorporate two-dimensional NMR experiments and complementary multi-modal spectroscopic data will be essential to resolve structural ambiguities that remain opaque to 1D signals. A specific challenge remains stereochemical differentiation, where the system’s accuracy currently lags behind its topology matching capabilities. To address this, future work will explore the integration of conformational ensemble modeling with computational chemistry methods to better capture spatial configurations. 

Ultimately, the implications of this work extend beyond NMR analysis to broader questions of cross-modal learning in the physical sciences. Our results demonstrate three key principles. First, effective integration demands synergistic coupling rather than simple concatenation of modalities. Second, learned representations and physics-based constraints serve complementary rather than redundant functions. Third, experimental training data remains essential for robust real-world performance. These principles provide a framework for AI-driven discovery across synthetic chemistry, pharmaceutical development, and biological systems, where computational approaches must reconcile the enduring divide between idealized theory and experimental practice.

\section{Conclusion}\label{Conclusion}

In this work, we present NMRPeak, a unified cross-modal learning framework that bridges the persistent discrepancy between computational simulation and experimental reality in NMR-based structure elucidation. We construct the largest curated benchmark of experimental NMR spectra to date and systematically quantify the distribution shift from the simulated data, providing a robust foundation for the field.
Our framework addresses two fundamental challenges in spectral analysis. The chemically-aware adaptive tokenizer resolves the trade-off between sparsity and semantic resolution through domain-informed discretization that aligns with experimental reporting standards. The assignment-free peak-aware similarity metric enables rigorous, physically grounded comparisons between predicted and experimental spectra without requiring atom-level assignments.
Crucially, NMRPeak demonstrates that integrating prediction, retrieval, and generation yields synergistic improvements. Forward prediction effectively bridges the simulation-to-experiment gap to enhance inverse inference. Through synergistic modular interaction, the system achieves over 95\% top-1 accuracy in spectrum-to-molecule retrieval and approximately 75\% top-1 accuracy in stereochemistry-aware \textit{de novo} structure generation on rigorous experimental benchmarks.
These results establish a new paradigm for AI-assisted NMR spectroscopy in which the field moves from isolated, simulation-dependent tasks to holistic, experimentally grounded systems capable of handling real-world complexity. This advance paves the way for fully automated, high-throughput molecular structure elucidation in chemical and biological discovery pipelines.

\section{Methods}\label{Methods}

\subsection{Data curation and benchmark construction}

\subsubsection{Benchmark datasets}

\paragraph{Experimental NMR dataset}
The experimental NMR dataset (NMRexp~\cite{wang2025nmrexp}) was curated through a multi-stage cleaning and harmonization pipeline designed to ensure chemical validity, spectral consistency, and compatibility with downstream prediction, retrieval, and generation tasks.

\textbf{Initial molecule validation and canonicalization.}
For each record, the associated molecular structure was first validated using RDKit~\cite{landrum2025rdkit}. Invalid or unparsable SMILES strings were discarded. All remaining molecules were standardized to a unique RDKit canonical SMILES representation, which served as the molecule-level identifier throughout data processing. Molecules failing sanitization, containing unsupported elements, exceeding 100 atoms, or lacking valid 3D coordinates were excluded. To reduce stereochemical ambiguity in experimental spectra, molecules with more than one chiral center were removed. Only molecules composed of elements from the set \{H, C, O, N, S, F, Cl, Br, Si, P, I, B, Se\} were retained. These filtering steps are necessitated by the performance limitations of optical chemical structure recognition (OCSR) tools~\cite{fang2025molparser, fang2025uni}, thereby prioritizing the accuracy of the resolved molecular structures.

\textbf{Independent cleaning of $^{13}$C NMR spectra.}
$^{13}$C NMR spectra were processed independently from $^{1}$H spectra due to their distinct annotation structure. Each spectrum was required to contain exclusively numeric chemical shift values within a physically plausible range (-50 to 300~ppm). Spectra containing malformed or non-numeric entries were discarded. To enforce internal chemical consistency, spectra were required to satisfy two constraints: (i) the number of reported $^{13}$C resonances did not exceed the number of carbon atoms in the molecular structure; and (ii) chemical shifts followed a monotonic ordering, consistent with standardized experimental reporting. Spectra violating either condition were removed.

\textbf{Independent cleaning of $^{1}$H NMR spectra.}
$^{1}$H NMR spectra were subjected to a more stringent validation procedure reflecting their richer annotation structure. Each peak was required to contain a valid multiplicity category, coupling constant annotations, proton count (integration), and a chemical shift range. Coupling constants were parsed numerically and constrained to physically plausible limits, with stricter upper bounds applied to molecules lacking heavy heteroatoms. Proton counts were required to be positive integers and capped to exclude implausible annotations. The chemical shift ranges were restricted to -3 to 17~ppm, and the malformed or inverted ranges were rejected. For each peak, the centroid of the reported chemical shift range was used as the representative chemical shift. To ensure molecular-level consistency, the sum of all integrated proton counts was required to exactly match the total number of hydrogen atoms in the molecular structure. Chemical shifts were additionally required to follow a monotonic ordering.

\textbf{Molecule-level merging and harmonization.}
After modality-specific cleaning, $^{1}$H and $^{13}$C spectra were merged at the molecule level using canonical SMILES as the unique key. For each molecule, at most one cleaned $^{1}$H spectrum and one cleaned $^{13}$C spectrum were retained. Molecules possessing only one of the two spectral modalities were preserved and included in the dataset. For each retained molecule, the molecular formula was calculated using RDKit and stored alongside spectral annotations, atomic identities, and 3D coordinates. Spectrometer frequency and solvent information, due to their highly concentrated and imbalanced distributions, are retained in the dataset but excluded from model training.

\textbf{Condition-aware sample filtering during training and evaluation.}
During training and validation, spectra were treated as valid samples only if all experimental settings required by the specific model configuration were available, including nucleus type ($^{1}$H and/or $^{13}$C). Records missing any required condition were filtered out under the corresponding experimental setting but remained available for other configurations. This design avoids imputing missing experimental metadata while maximizing data reuse across different learning tasks.

\paragraph{Simulated NMR dataset}

The preprocessing of MST-NMR~\cite{alberts2024unraveling} focused primarily on molecular validity rather than spectral consistency. For each record, the associated molecular structure was validated using RDKit, and invalid or unparsable SMILES strings were discarded. All remaining molecules were standardized to RDKit canonical SMILES representations to ensure consistency with the experimental dataset.

No additional spectrum-specific filtering was applied to MST-NMR, as the simulated $^{1}$H and $^{13}$C spectra are noise-free, physically consistent by construction, and fully paired with the corresponding molecular structures. This distinction allows MST-NMR to serve as a clean simulation baseline while preserving a clear separation from the more heterogeneous experimental NMR data.

\subsubsection{Data splitting}

Although MST-NMR provided the raw dataset, the split datasets were not available. Therefore, we re-curated both datasets by following the same splitting strategy, which was consistently applied across experimental (NMRexp) and simulated (MST-NMR) datasets.

All datasets were split into training, validation, and test subsets following a randomized molecule-level protocol. For each dataset, samples were first randomly partitioned into a training set (90\%) and a held-out test set (10\%). The training set was then further split into a reduced training subset (95\%) and a validation subset (5\%). This procedure resulted in final proportions of 85.5\%, 4.5\%, and 10\% for training, validation, and test sets, respectively. All splits were performed using a fixed random seed to ensure reproducibility. Random shuffling was applied prior to splitting, and each molecule was assigned to exactly one subset, preventing any overlap between training, validation, and test data. 

\subsection{Chemically-aware adaptive tokenizer}

We introduce a chemically-aware adaptive tokenizer that converts heterogeneous spectral annotations and molecular formula information into a unified discrete sequence. The tokenizer is designed to balance numerical precision, chemical interpretability, and vocabulary efficiency, while remaining robust to experimental noise.

\subsubsection{Token categories and vocabulary design}

The overall vocabulary consists of three classes of tokens: special tokens, categorical tokens, and numerical tokens (Fig.~\ref{fig:framework}b).

\textbf{Special tokens} include sequence boundary markers (\texttt{BOS} and \texttt{EOS}) as well as modality-specific separators indicating the start and end of $^{13}$C NMR spectra and $^{1}$H NMR spectra segments. These tokens enable a single sequence to encode multiple modalities in a fixed and interpretable order.

\textbf{Categorical tokens} represent discrete chemical attributes that are naturally symbolic rather than numerical. For $^{1}$H NMR spectra, this includes peak multiplicity categories and proton integration counts. For molecular formulae, categorical tokens encode element types and corresponding atom counts.

\textbf{Numerical tokens} are used to represent continuous-valued spectral quantities, including $^{13}$C chemical shifts, $^{1}$H chemical shifts, and $^{1}$H coupling constants. These variables are discretized using an adaptive binning strategy described below.

\subsubsection{Adaptive discretization of numerical variables}

Direct uniform discretization of spectral values either leads to excessively large vocabularies when using fine bins or to severe information loss when using coarse bins. To address this trade-off, we adopt a chemically-aware adaptive discretization strategy that assigns different resolutions to different value ranges based on empirical peak density and chemical relevance.

For each numerical variable, the value range is partitioned into a set of non-overlapping intervals. Within high-density or chemically informative regions, finer bin widths are used to preserve resolution, whereas coarser bins are applied in low-density or less informative regions. Extreme values outside typical experimental ranges are grouped into overflow bins.

Each numerical token corresponds to a closed interval and is represented by a unique symbolic identifier encoding its lower and upper bounds. The overall vocabulary size is thus controlled by the number of adaptive intervals rather than by the numerical precision of the original values.

\subsubsection{Spectrum sequence construction and inverse transformation}

Each NMR spectrum is converted into a sequence of tokens by iterating over individual peaks. For $^{13}$C NMR spectra, each peak contributes a single numerical token representing its chemical shift. For $^{1}$H NMR spectra, each peak is encoded as a composite of categorical and numerical tokens, including multiplicity, coupling constant tokens, proton count, and chemical shift (range).

Peaks within each spectrum are ordered according to decreasing chemical shift, consistent with standard experimental reporting conventions. $^{13}$C and $^{1}$H spectra are concatenated into a single sequence using modality-specific separator tokens, followed by a molecular formula segment encoding element types and counts.

Variable-length sequences are supported by the use of explicit boundary tokens and standard padding during batching. Each numerical token maps to a predefined value interval, and inverse transformation yields the interval midpoint as the reconstructed value. Consequently, NMRPeak-P predictions may exhibit discrete clustering where multiple distinct experimental values are mapped to identical predicted values, manifesting as horizontal striping patterns in the correlation plots (see Fig.~\ref{fig:prediction_results}b).

\subsection{Peak-aware similarity metric}

\subsubsection{Problem formulation}
Given two NMR spectra of the same type (either $^1$H NMR or $^{13}$C NMR), we aim to compute a normalized similarity score in the range $[0, 1]$ that quantifies their structural resemblance. Each spectrum is represented as an unordered set of peaks without atom-level assignments. For $^1$H NMR, each peak is characterized by a five-tuple: chemical shift range $[\delta_{\min}, \delta_{\max}]$, multiplicity, integration (number of hydrogens $n_{\text{H}}$), and coupling constants. For $^{13}$C NMR, each peak is represented solely by its chemical shift $\delta$. The similarity score enables direct evaluation of NMRPeak-P predictions and facilitates spectrum-to-molecule re-ranking in both NMRPeak-R and NMRPeak-G pipelines.

Formally, let $\mathcal{A} = \{a_1, a_2, \ldots, a_{n_A}\}$ and $\mathcal{B} = \{b_1, b_2, \ldots, b_{n_B}\}$ denote two spectra with $n_A$ and $n_B$ peaks, respectively. Our objective is to define a similarity function $S(\mathcal{A}, \mathcal{B}) \in [0, 1]$ that captures both the quality of peak-to-peak correspondences and the overall consistency between the two spectra.

\subsubsection{Two-stage peak matching procedure}
The peak matching process consists of two stages designed to establish comprehensive correspondences between peaks in $\mathcal{A}$ and $\mathcal{B}$.

We first define a distance-based similarity function that quantifies the proximity between spectral features:
\begin{equation}
\phi(d, \sigma) = \exp\left(-\frac{d}{\sigma}\right)
\end{equation}
where $d$ is the distance between two features (e.g., chemical shift difference) and $\sigma$ is the scale parameter that controls the decay rate of similarity with distance. We use $\sigma = 5.0$ ppm for $^{13}$C NMR and $\sigma = 1.0$ ppm for $^1$H NMR as default values. Additionally, we define a tolerance threshold $\tau$ beyond which matches are considered invalid: $\tau = 20.0$ ppm for $^{13}$C NMR and $\tau = 2.0$ ppm for $^1$H NMR.

\textbf{Peak expansion for $^1$H NMR.}
For $^1$H NMR spectra, we employ an expansion strategy to account for the fact that each peak represents multiple hydrogens. Each peak $a_i$ with integration $n_{\text{H}}^{(i)}$ is replicated $n_{\text{H}}^{(i)}$ times to create virtual peaks with identical chemical shifts. After expansion, let $\mathcal{A}'$ and $\mathcal{B}'$ denote the expanded peak sets with sizes $n'_A = \sum_{i=1}^{n_A} n_{\text{H}}^{(i)}$ and $n'_B = \sum_{j=1}^{n_B} n_{\text{H}}^{(j)}$, respectively. For $^{13}$C NMR, no expansion is performed, so $\mathcal{A}' = \mathcal{A}$ and $\mathcal{B}' = \mathcal{B}$.

\textbf{Stage I: Global optimal matching.} 
Without loss of generality, assume $|\mathcal{A}'| \leq |\mathcal{B}'|$. We construct a cost matrix $\mathbf{C} \in \mathds{R}^{|\mathcal{A}'| \times |\mathcal{B}'|}$ for the linear assignment problem, with different formulations for $^{13}$C and $^1$H NMR.

For $^{13}$C NMR, we first compute a similarity matrix:
\begin{equation}
S_{ij} = \phi(|\delta_{a_i} - \delta_{b_j}|, \sigma)
\end{equation}
and then convert it to a cost matrix:
\begin{equation}
C_{ij} = 1 - S_{ij}
\end{equation}

For $^1$H NMR (using expanded peaks), we directly construct a squared-distance cost matrix based on the center of chemical shift ranges:
\begin{equation}
C_{ij} = (\delta_{a_i}^{\text{center}} - \delta_{b_j}^{\text{center}})^2
\end{equation}
where $\delta^{\text{center}} = (\delta^{\min} + \delta^{\max})/2$.

We solve the linear assignment problem using the Hungarian algorithm to find the optimal one-to-one matching $\mathcal{M}_1 = \{(i, \pi(i)) : i = 1, \ldots, |\mathcal{A}'|\}$, where $\pi$ is the assignment function that minimizes the total cost:
\begin{equation}
\pi^* = \arg\min_{\pi} \sum_{i=1}^{|\mathcal{A}'|} C_{i,\pi(i)}
\end{equation}

\textbf{Stage II: Greedy matching of remaining peaks.}
After Stage I, the larger spectrum $\mathcal{B}'$ contains $|\mathcal{B}'| - |\mathcal{A}'|$ unmatched peaks. For each remaining peak $b_j \in \mathcal{B}'$ that was not matched in Stage I, we perform a greedy search to find its best match in $\mathcal{A}'$ by maximizing the similarity score.

For $^{13}$C NMR:
\begin{equation}
i^* = \arg\max_{i \in \{1, \ldots, |\mathcal{A}'|\}} \phi(|\delta_{a_i} - \delta_{b_j}|, \sigma)
\end{equation}

For $^1$H NMR:
\begin{equation}
i^* = \arg\max_{i \in \{1, \ldots, |\mathcal{A}'|\}} \phi(|\delta_{a_i}^{\text{center}} - \delta_{b_j}^{\text{center}}|, \sigma)
\end{equation}

Let $d^* = |\delta_{a_{i^*}} - \delta_{b_j}|$ for $^{13}$C or $d^* = |\delta_{a_{i^*}}^{\text{center}} - \delta_{b_j}^{\text{center}}|$ for $^1$H. The match $(i^*, j)$ is added to $\mathcal{M}_2$ and is considered valid if $d^* \leq \tau$; otherwise, it is marked as invalid. This greedy strategy ensures all peaks are matched while maintaining quality assessment.

\subsubsection{Similarity scoring with multi-feature integration}
The final similarity score combines the averaged peak-level similarities with penalty terms that account for spectral inconsistencies.

\textbf{Peak-level similarity for $^{13}$C NMR.}
For each matched pair $(i, j) \in \mathcal{M}_1 \cup \mathcal{M}_2$, the similarity depends solely on chemical shift:
\begin{equation}
s_{ij}^{\text{C}} = \begin{cases}
\phi(|\delta_{a_i} - \delta_{b_j}|, \sigma) & \text{if } |\delta_{a_i} - \delta_{b_j}| \leq \tau \\
0 & \text{otherwise}
\end{cases}
\end{equation}

Matches from Stage II are dampened by a factor of 0.8 to reflect their greedy nature:
\begin{equation}
\tilde{s}_{ij}^{\text{C}} = \begin{cases}
s_{ij}^{\text{C}} & \text{if } (i,j) \in \mathcal{M}_1 \\
0.8 \cdot s_{ij}^{\text{C}} & \text{if } (i,j) \in \mathcal{M}_2
\end{cases}
\end{equation}

The average peak similarity is:
\begin{equation}
\bar{s}^{\text{C}} = \frac{1}{|\mathcal{M}_1 \cup \mathcal{M}_2|} \sum_{(i,j) \in \mathcal{M}_1 \cup \mathcal{M}_2} \tilde{s}_{ij}^{\text{C}}
\end{equation}

The structural penalty accounts for peak count differences and invalid matches:
\begin{equation}
p_{\text{struct}}^{\text{C}} = \frac{\max(|\mathcal{A}|, |\mathcal{B}|)}{|\mathcal{M}_1 \cup \mathcal{M}_2| + ||\mathcal{A}| - |\mathcal{B}|| + (|\mathcal{M}_1 \cup \mathcal{M}_2| - |\mathcal{M}_1^{\text{valid}}|)}
\end{equation}
where $\mathcal{M}_1^{\text{valid}}$ denotes the subset of Stage I matches that satisfy the distance threshold.

The final $^{13}$C NMR similarity score is:
\begin{equation}
S(\mathcal{A}, \mathcal{B}) = 0.8 \cdot \bar{s}^{\text{C}} + 0.2 \cdot p_{\text{struct}}^{\text{C}}
\end{equation}

\textbf{Peak-level similarity for $^1$H NMR.}
For $^1$H NMR, we integrate multiple spectral features with weighted contributions. For each matched pair of expanded peaks $(i, j)$ that map back to original peaks $(a_{\mu(i)}, b_{\nu(j)})$ through the expansion mapping, we compute:
\begin{align}
s_{ij}^{\text{H}} = &\ 0.4 \cdot \phi(|\delta_{\mu(i)}^{\min} - \delta_{\nu(j)}^{\min}|, \sigma) + 0.4 \cdot \phi(|\delta_{\mu(i)}^{\max} - \delta_{\nu(j)}^{\max}|, \sigma) \nonumber \\
&\ + 0.05 \cdot \mathds{1}[\text{mult}(a_{\mu(i)}) = \text{mult}(b_{\nu(j)})] \nonumber \\
&\ + 0.1 \cdot \mathds{1}[n_{\text{H}}(a_{\mu(i)}) = n_{\text{H}}(b_{\nu(j)})] \nonumber \\
&\ + 0.05 \cdot s_{\text{coupling}}(a_{\mu(i)}, b_{\nu(j)})
\end{align}
where $\mathds{1}[\cdot]$ is the indicator function, $\text{mult}(\cdot)$ denotes peak multiplicity, and $s_{\text{coupling}}$ computes the similarity between sorted coupling constant lists:
\begin{equation}
s_{\text{coupling}}(a, b) = \begin{cases}
\frac{1}{|J_a|} \sum_{k=1}^{|J_a|} \phi(|J_a^{(k)} - J_b^{(k)}|, 10) & \text{if } |J_a| = |J_b| \\
0 & \text{otherwise}
\end{cases}
\end{equation}
where $J_a$ and $J_b$ are the sorted coupling constant lists (in Hz), and the scale parameter 10 Hz is used for coupling constant comparison.

The validity check and Stage II dampening are applied as:
\begin{equation}
\tilde{s}_{ij}^{\text{H}} = \begin{cases}
s_{ij}^{\text{H}} & \text{if } (i,j) \in \mathcal{M}_1 \text{ and } d_{ij} \leq \tau \\
0.8 \cdot s_{ij}^{\text{H}} & \text{if } (i,j) \in \mathcal{M}_2 \text{ and } d_{ij} \leq \tau \\
0 & \text{otherwise}
\end{cases}
\end{equation}
where $d_{ij} = |\delta_{\mu(i)}^{\text{center}} - \delta_{\nu(j)}^{\text{center}}|$.

The average peak similarity is:
\begin{equation}
\bar{s}^{\text{H}} = \frac{1}{|\mathcal{M}_1 \cup \mathcal{M}_2|} \sum_{(i,j) \in \mathcal{M}_1 \cup \mathcal{M}_2} \tilde{s}_{ij}^{\text{H}}
\end{equation}

For $^1$H NMR, we compute penalties based on the expanded peak sets:
\begin{equation}
p_{\text{struct}}^{\text{H}} = \frac{\max(|\mathcal{A}'|, |\mathcal{B}'|)}{|\mathcal{M}_1 \cup \mathcal{M}_2| + ||\mathcal{A}'| - |\mathcal{B}'|| + (|\mathcal{M}_1 \cup \mathcal{M}_2| - |\mathcal{M}_1^{\text{valid}}|)}
\end{equation}

The hydrogen count penalty reflects the overall hydrogen balance:
\begin{equation}
p_{\text{H}} = \frac{\min(|\mathcal{A}'|, |\mathcal{B}'|)}{\max(|\mathcal{A}'|, |\mathcal{B}'|)}
\end{equation}

The final $^1$H NMR similarity score is:
\begin{equation}
S(\mathcal{A}, \mathcal{B}) = (0.8 \cdot \bar{s}^{\text{H}} + 0.2 \cdot p_{\text{struct}}^{\text{H}}) \cdot p_{\text{H}}
\end{equation}








\subsection{Training procedure}

\subsubsection{Learning objectives}

We train three distinct models with different objectives tailored to their respective tasks: spectrum prediction (NMRPeak-P), spectrum-molecule retrieval (NMRPeak-R), and molecule generation (NMRPeak-G).

\textbf{Spectrum prediction loss of NMRPeak-P.}
NMRPeak-P is trained to autoregressively predict spectral tokens from molecular structure. The model takes as input molecular features encoded using the same protocol as Uni-Mol~\cite{zhouuni} (atomic elements and 3D coordinates when available), and generates a sequence of spectral tokens obtained from the chemistry-aware tokenizer. The model is trained with a standard autoregressive language modeling objective:

\begin{equation}
\mathcal{L}_{\text{prediction}} = -\sum_{t=1}^{T} \log P(z_t | z_{<t}, \mathbf{m}; \theta)
\end{equation}

where $\mathbf{m}$ denotes the molecular representation, $z_t$ is the $t$-th spectral token in the sequence of length $T$, and $\theta$ represents the model parameters. The predicted spectral tokens are subsequently decoded back to the original peak list representation (chemical shifts, multiplicities, integrations, and coupling constants) through inverse transformation of the tokenizer.

\textbf{Contrastive retrieval loss of NMRPeak-R.}
NMRPeak-R learns aligned representations of molecules and spectra through contrastive learning. The model encodes molecular features $\mathbf{m}$ into embedding $\mathbf{e}_{\text{mol}}$ and spectral tokens $\mathbf{z}$ (from the chemistry-aware tokenizer) into embedding $\mathbf{e}_{\text{spec}}$. For a batch of $B$ molecule-spectrum pairs, the model computes a similarity matrix scaled by temperature $\tau_{\text{cl}}$:

\begin{equation}
\mathbf{S} = \frac{\mathbf{E}_{\text{mol}} \mathbf{E}_{\text{spec}}^T}{\tau_{\text{cl}}}
\end{equation}

where $\mathbf{E}_{\text{mol}} \in \mathds{R}^{B \times d}$ and $\mathbf{E}_{\text{spec}} \in \mathds{R}^{B \times d}$ are the normalized embedding matrices. The contrastive loss is computed bidirectionally:

\begin{align}
\mathcal{L}_{\text{mol}\rightarrow\text{spec}} &= -\frac{1}{B}\sum_{i=1}^{B} \log \frac{\exp(S_{ii})}{\sum_{j=1}^{B} \exp(S_{ij})} \\
\mathcal{L}_{\text{spec}\rightarrow\text{mol}} &= -\frac{1}{B}\sum_{i=1}^{B} \log \frac{\exp(S_{ii})}{\sum_{j=1}^{B} \exp(S_{ji})} \\
\mathcal{L}_{\text{retrieval}} &= \frac{\mathcal{L}_{\text{mol}\rightarrow\text{spec}} + \mathcal{L}_{\text{spec}\rightarrow\text{mol}}}{2}
\end{align}

where the diagonal entries $S_{ii}$ correspond to positive pairs (matched molecule-spectrum pairs) and off-diagonal entries represent negative pairs.

Notably, molecular formula information is \textit{not} explicitly incorporated as input tokens during NMRPeak-R training. This design choice is motivated by two considerations: (1) In retrieval scenarios where the search space is constrained to a finite database, molecular formula filtering as a post-processing step already achieves high precision without requiring the model to learn this association. (2) Providing formula information during training may inadvertently cause the model to rely on this high-level constraint rather than learning fine-grained spectral features, potentially compromising its ability to distinguish between isomers with identical formulas.

\textbf{Molecular generation loss of NMRPeak-G.}
NMRPeak-G performs conditional molecule generation from spectral information. The model takes as input spectral tokens $\mathbf{z}$ and optionally molecular formula tokens $\mathbf{f}$ from the chemistry-aware adaptive tokenizer, and autoregressively generates SMILES tokens $\{s_1, s_2, \ldots, s_L\}$ obtained from Smirk-based tokenization~\cite{wadell2024smirk}:

\begin{equation}
\mathcal{L}_{\text{generation}} = -\sum_{t=1}^{L} \log P(s_t | s_{<t}, \mathbf{z}, \mathbf{f}; \theta)
\end{equation}

where $s_t$ is the $t$-th SMILES token in the sequence of length $L$ and $\theta$ are the model parameters.

In contrast to NMRPeak-R, molecular formula information must be \textit{explicitly provided} as input tokens when available. This arises from the fact that the generation search space becomes unbounded without molecular formula constraints, the model would need to explore an intractably large space of possible molecules. By conditioning on the molecular formula, we effectively reduce the search space to molecules satisfying the elemental composition, thereby improving both generation efficiency and accuracy. When formula information is unavailable, the model generates molecules conditioned solely on spectral tokens.

\subsubsection{Model architecture}

The overall framework is built upon two backbone architectures: a Uni-Mol~\cite{zhouuni} encoder for molecular representation learning and a BART-based~\cite{lewis2020bart} encoder--decoder architecture for SMILES sequence and spectral sequence modeling. Depending on the task, different combinations are employed while maintaining a unified design philosophy.

For all modules involving molecular structure encoding, we adopt the Uni-Mol encoder architecture~\cite{zhouuni}. The input consists of atomic element types and 3D coordinates, following the original Uni-Mol preprocessing protocol. The pretrained Uni-Mol weights are initialized from the official checkpoint but are not frozen; instead, the entire encoder is fine-tuned during training to adapt to NMR-related tasks. For modules involving molecular sequence generation, SMILES tokens obtained via Smirk-based tokenization~\cite{wadell2024smirk} are processed using a BART decoder. The decoder follows the standard BART decoder configuration and autoregressively generates SMILES tokens conditioned on encoder outputs. Spectral tokens produced by the chemistry-aware adaptive tokenizer are processed using a standard BART architecture. Specifically, the spectral encoder and spectral decoder both follow the BART-base configuration. All BART-based components share the same vocabulary. When encoder and decoder are both from BART, their token embedding layers are shared, consistent with the original BART design.

\textbf{Architecture of NMRPeak-P.}
NMRPeak-P adopts a molecule-to-spectrum encoder--decoder architecture. The Uni-Mol encoder encodes molecular features into continuous representations, which are then fed into a BART decoder that autoregressively generates spectral tokens discretized by the chemistry-aware tokenizer. 

\textbf{Architecture of NMRPeak-R.} 
NMRPeak-R employs a dual-encoder architecture for contrastive learning. Molecular structures are encoded by the Uni-Mol encoder, while spectral tokens are encoded by a BART encoder. To align the representation spaces, a fully connected projection layer (adapter) is appended to the Uni-Mol encoder output, projecting molecular embeddings to the same dimensionality as the spectral encoder (768 dimensions). The resulting molecular and spectral embeddings are $\ell_2$-normalized before computing the similarity matrix for contrastive training. In this module, no decoder is used.

\textbf{Architecture of NMRPeak-G.}
NMRPeak-G follows a spectrum-to-molecule encoder--decoder architecture. Spectral tokens (and optionally molecular formula tokens) are first encoded by a BART encoder. The encoded spectral representations are then consumed by a BART decoder that autoregressively generates SMILES tokens using Smirk-based tokenization~\cite{wadell2024smirk}. Since both encoder and decoder follow the standard BART configuration, they share token embedding parameters. When molecular formula tokens are provided, they are concatenated with spectral tokens at the encoder input level, enabling the decoder to condition generation on both spectral patterns and elemental composition constraints.

\subsubsection{Multi-model ensemble strategy for NMRPeak-P}

For inference with a single-model version of NMRPeak-P, we use the best-performing checkpoint selected according to the validation performance. For the multi-model ensemble version, we leverage multiple checkpoints saved at different training steps. Instead of relying on a single best model, predictions are generated from several intermediate checkpoints and then aggregated during inference. By default, we ensemble 10 models selected from different training steps. 

\subsubsection{Retrieval weight calibration for NMRPeak-R}
The combined score is computed as a linear combination of SME, SSE, and SSR:
\begin{equation}
S_{\text{combine}} = \alpha \cdot \text{SME} + \beta \cdot \text{SSE} + \gamma \cdot \text{SSR},
\end{equation}
where $\alpha + \beta + \gamma = 1$. Each coefficient takes values in the range $[0, 1]$ with an interval of 0.1. Therefore, for each fixed input condition, there are 66 possible coefficient combinations.

When the input simultaneously contains both $^{13}$C and $^1$H spectra, SME and SSE are computed directly on the combined input, yielding a single score for each metric. In contrast, the SSR score is calculated as a weighted combination of the two modalities:
\begin{equation}
\text{SSR} = \lambda \cdot \text{SSR}_{^{13}\mathrm{C}} + (1 - \lambda) \cdot \text{SSR}_{^1\mathrm{H}},
\end{equation}
where $\lambda \in [0,1]$ with a step size of 0.1, and the two weights sum to 1.

\subsubsection{Beam search strategy for NMRPeak-G}

The beam search implementation follows the decoding strategies used in XLM models~\cite{conneau2019cross} and the NAG2G model~\cite{yao2024node}. During inference, we adopt task-specific decoding hyperparameters to balance determinism and diversity. For NMRPeak-P, the default beam size is set to 1 and the temperature is set to 1.0, resulting in deterministic and stable spectral predictions. For NMRPeak-G, the default beam size is set to 10 and the temperature is set to 3.0, encouraging higher diversity in generated molecular structures. 

\subsubsection{Computational resources}

All models were trained on NVIDIA A100 GPUs with 80GB memory. 
NMRPeak-P and NMRPeak-G were trained on 8 $\times$ A100 80GB GPUs using distributed data parallel training, whereas NMRPeak-R was trained on a single A100 80GB GPU.

\subsection{Evaluation protocol}

The evaluation of NMRPeak-P is conducted from two complementary perspectives. First, we compute the similarity score using the peak-aware similarity metric, which directly measures the agreement between predicted and reference spectra at the peak level (including chemical shifts, multiplicities, integrations, and coupling constants). Notably, this metric is highly stringent, and even minor deviations in peak attributes can lead to noticeable score reductions. As a result, similarity scores are typically below 0.95, even for high-quality predictions. Second, we assess the structural consistency of predicted spectra by feeding them into a frozen NMRPeak-G model to reconstruct molecular structures. We report top-$k$ accuracy, measuring whether the ground-truth molecule appears among the top-$k$ generated candidates. This evaluation reflects the practical utility of predicted spectra in downstream structure elucidation. Due to the strictness of the peak-aware similarity metric, even small numerical differences between NMRPeak-P-Single and NMRPeak-P-Multi correspond to substantial differences in actual spectral parsing accuracy. 

For NMRPeak-R and NMRPeak-G, the evaluation metrics are based on top-$k$ retrieval and generation accuracy, respectively. Notably, the retrieval accuracy of NMRPeak-R is reported over the entire test set as the candidate search space.












\backmatter

\bmhead{Data availability}
All processed NMR datasets used in the NMRPeak benchmark are available at \url{https://zenodo.org/records/19122815}. The dataset includes all curated data splits of NMRexp~\cite{wang2025nmrexp} and MST-NMR~\cite{alberts2024unraveling} used for training, validation, and testing, along with the original raw data from NMRexp.

\bmhead{Code availability}
The NMRPeak code is publicly available at \url{https://github.com/Colin-Jay/NMRPeak}, and the trained model parameters are available at \url{https://zenodo.org/records/19122815}. An online service is also available at \url{https://ai4ec.ac.cn/apps/nmrpeak}.





\bmhead{Acknowledgements}

F.T. acknowledges the National Key R\&D Program of China (Grant No. 2024YFA1210804), National Natural Science Foundation of China (Grant No. 22573085), and a startup fund at Xiamen University. J.C. acknowledges the National Natural Science Foundation of China (Grant Nos. 22225302, 92470201, 92461312, 22021001, 21991151, 21991150, 92161113, 22411560277), the Fundamental Research Funds for the Central Universities (20720220009, 20720250005), and the Laboratory of AI for Electrochemistry (AI4EC) and IKKEM (Grant Nos. RD2023100101 and RD2022070501) for financial support. This work used the computational resources in the IKKEM intelligent computing center. J.Z. acknowledges the National Natural Science Foundation of China (Grants No. 22403037), Laboratory of AI for Electrochemistry (AI4EC) and IKKEM (Grants Nos. RD2023100101 and RD2022070501) for financial support.

\bibliography{sn-bibliography}

\end{document}